\documentclass[preprint,showkeys,aip]{revtex4-1}
\usepackage{amsmath}
\usepackage{latexsym}
\usepackage{amssymb}
\usepackage{graphics,epstopdf}
\usepackage{amsmath,amssymb,amsthm}
\usepackage{mathrsfs}
\usepackage{graphicx}
\usepackage[colorlinks=true, citecolor=blue, urlcolor=blue ]{hyperref}
\usepackage{cleveref}

\begin{document}
	\title{Estimation of covariance matrix for Stokes-like measurement under phase-space deformation}
	\author{Pinaki Patra}
	\thanks{Corresponding author}
	\email{monk.ju@gmail.com}
	\affiliation{Department of Physics, Brahmananda Keshab Chandra College, 111/2 B. T. Road, Kolkata, India-700108}

	\date{\today}

	\begin{abstract}
	We investigate the signatures of phase-space noncommutativity in an anisotropic oscillator and propose an optomechanical protocol to experimentally characterize it through Stokes-like measurements of an optical field. Starting from a deformed phase-space algebra characterized by the position-position and momentum-momentum noncommutativity parameters $\theta$ and $\eta$, we employ the Darboux transformation to construct an equivalent Hamiltonian in canonical variables. The resulting effective dynamics exhibits correlations between the oscillator degrees of freedom. We derive the corresponding covariance matrix and examine how the noncommutative parameters enter the mechanical dynamics and the optical response. \\
	By coupling the oscillator to a coherent optical probe, we obtain analytical expressions for the time-dependent optical coherence attenuation and the nonlinear Kerr phase. We then formulate an interferometric detection scheme through Stokes-like measurements, in which the outgoing optical field is combined with a phase-controlled reference beam at a balanced beam splitter, and the resulting photocurrents are measured using photodetectors. The optical response provides access to the modified normal-mode frequencies, which enables us to estimate experimental values of  $\theta$ and $\eta$.  Our results establish a theoretical connection between NC-space dynamics, Gaussian-state correlations, and experimentally accessible optical observables, suggesting an optomechanical route toward testing signatures of phase-space noncommutativity.
	\end{abstract}

		
	\maketitle
	
	\section{Introduction}
	Quantum entanglement (verschr\"{a}nkung) implies the existence of global states of a composite system which can not be written as a product of the states of individual
	subsystems \cite{entanglement1}. It underlines the intrinsic order of statistical correlations between subsystems of a compound quantum system \cite{entanglement2}. The insight into quantum entanglement leads to an understanding of a diverse range of phenomena in which correlations are important \cite{decoherence}. For instance, the understanding of quantum optics, phase transitions in condensed matter systems, black hole physics, and the relation between the geometrical structure of the dual spacetime and entanglement structure of the conformal field theory have become topical issues \cite{entanglement3,entanglement4,entanglement5,entanglement6}. One of the key goals of modern physics is to understand the physical origin(s) of entanglement, which is still obscure \cite{entorigin,classmed1}. 
   Besides quantum mechanics (QM), we have another cornerstone of modern physics, namely Gravity, which describes the dynamics of spacetime itself \cite{rovelli1}. There is no single reproducible reported violation of QM and gravity to date. We expect that QM and gravity are subsets of some unified theory, namely quantum gravity (QG). There are many elegant, consistent theoretical proposals of QG, namely string theory, loop quantum gravity, M-theory, twistor theory, and so on \cite{quantumgravity1,quantumgravity2,quantumgravity3,quantumgravity4}. However, the lack of experimental confirmation of any of the proposals of QG raises serious doubts about whether gravity falls under the premise of quantum theory at all \cite{hanif1}. Most probably, the lack of successful experiments for QG is not due to the drawbacks of the proposed theories. Rather, it is due to the limitation of present-day experimental capacity. By way of comparison, the LHC was designed to run at a maximum collision energy of $14$  TeV \cite{lhc}, whereas most of the theories of quantum gravity appear to predict departures from classical gravity only at energy scales on the order of $10^{19}$ GeV \cite{qg1,qg2}. We hope that the high-energy scattering experiment may lead us to the feasibility of probing such high-energy scales in the future. However, colliders are costly. Naturally, alternative directions toward the exploration of the quantum signature of gravity are always fascinating. For instance,  proposals toward low-energy laboratory tests that can witness the nonclassicality of the gravitational field, with the help of conventional interferometric techniques, are available in the literature \cite{expt1,expt2,expt3,lightmatter1,lightmatter2,lightmatter3}. 
   It seems likely that noncommutative (NC) deformation to the symplectic structure of phase space, as brought about by position-position noncommutativity and momentum-momentum noncommutativity at a fundamental scale, could impact the correlations of coordinate and momentum degrees of freedom \cite{ncs3,nandi1,Moyal2,ncssim1,Gouba1,NCent1}. In particular, we expect a deep-rooted connection between NC dynamics and the phenomenon of entanglement of quantum states \cite{NCent2,NCent3,NCent4}. 
    \\
   In the line of investigation towards the proposals of experimental tests for QG, one often utilizes the consensus among most of the theoretical proposals that the quantization of spacetime itself can have testable experimental implications: the existence of a minimal length scale is expected to result in a modification of the Heisenberg uncertainty relation. In the present article, we aim to illustrate a theoretical proposal towards estimation of phase-space deformation parameters by experimental means through optomechanical scheme. In particular,  we utilize Stokes-like measurement schemes for quantum state tomography, in which a reference beam $\hat{a}_R$ is interfered with the output signal $\hat{a}_{L,out}$ that created through the optomechanical interactions inside an optical cavity.  The resultant beams are detected on photodetectors with measurements of photocurrents, which are proportional to photon numbers.   Data of average photocurrents provide the estimation of the expectation values of quadratures for various phases of input optical modes, thus the covariance matrix \cite{stokes1,stokes2,stokes3}. If the experimental set up is maintained at sufficiently low temperature (below $4$ Kelvin), the oscillator may be well approximated by its ground state, namely a Gaussian state, which is completely characterized by its first moment and covariance matrix.  If the estimated covariance matrix indicates entanglement between phase-space co-ordiantes, then it might be signature of NC-deformation of phase-space.  Moreover, we show that the estimation of the various quadratures through photocurrents measurement also provide experimental estimation of optical coherence attenuation $A_M$ and the nonlinear Kerr phase ($\Phi_K$) that produced via the interaction of the anisotropic oscilaltor and the input signal inside the optomechanical cavity. We explicitly show that the estimation of $A_M$ and $\Phi_K$ determines the normal mode frequency of the anisotropic oscillator by two frequency spectral estimation problem \cite{prony1,prony2}. Finally, we outlined how these experimental estimations determine the NC-phase space deformation parameters $\theta$ and $\eta$.\\
   However, it is important to distinguish the nonseparability of classical fields from the entanglement of quantum states. Classical systems possessing multiple degrees of freedom can exhibit nonfactorizable correlation structures analogous to those encountered in composite quantum systems. In particular, classical electromagnetic fields with coupled spatial and polarization degrees of freedom provide examples of such nonseparability, and certain associated correlations can reproduce mathematical features of Bell-type inequalities \cite{classent1,classent2,classent3}. In the present investigation, the term quantum correlation refers specifically to correlations arising within the quantum-mechanical description of a bipartite oscillator governed by a deformed phase-space algebra. Our objective is not to establish nonclassicality solely from the existence of correlations, but to examine how the assumed noncommutative structure modifies the Hamiltonian dynamics, Gaussian ground-state properties, and optomechanical response. 
   Moreover,  we should keep in mind that the generality of NC-geometry is not an
   easy question in QG (for instance, for closed strings in string theories)\cite{stringnc1,stringnc2}. In particular, NC phase space is not a universal prediction of quantum gravity. It is a particular mathematical framework that appears in certain approaches to quantum spacetime, but it is neither a necessary nor a sufficient condition for the quantization of gravity. Our proposed Stokes-like measurement scheme investigates whether deviations from the canonical phase-space algebra, if present and adequately described by the adopted deformation, can produce measurable modifications of the mechanical spectrum and the optical response.
   	\\
	The organization of the paper is as follows. At first, we discuss the entanglement generated by the noncommutative deformation of phase space for an anisotropic harmonic oscillator. Then we introduce the Stokes-like measurement scheme for an  oscillator state in noncommutative space, and outlined how NC-parameters be determined from the estimation of covariance.
\section{Entanglement generation by non-commutative deformations}
An anisotropic two dimensional harmonic oscillator in noncommutative space is described by the Hamiltonian 
\begin{equation}\label{NCShamiltonian}
	\hat{H}_{\mathrm{nc}}=\frac{1}{2} \hat{\tilde{X}}^T \mathcal{H}_{\mathrm{nc}}\hat{\tilde{X}},\; \mbox{with}\; \hat{\tilde{X}}= (\hat{\tilde{X}}_1,\hat{\tilde{X}}_2, \hat{\tilde{X}}_3, \hat{\tilde{X}}_4)^T= (\hat{\tilde{x}}_1,\hat{\tilde{p}}_1, \hat{\tilde{x}}_2, \hat{\tilde{p}}_2)^T.
\end{equation}
Here,  $X^T$ stands for matrix transposition of $X$. The anisotropic oscillator in the NC-space is characterized by   the mass $M=(m_1,m_2)$, and the frequency $\tilde{\omega}=(\tilde{\omega}_1,\tilde{\omega}_2)$ through
\begin{eqnarray}
\mathcal{H}_{\mathrm{nc}}= \left(\begin{array}{cc}
	\mathcal{H}_{\mathrm{nc}}^{(1)} & 0 \\
	0 & \mathcal{H}_{\mathrm{nc}}^{(2)}
\end{array}\right),\;
\mbox{with}\;
\mathcal{H}_{\mathrm{nc}}^{(j)}=\mbox{Diag}(m_j\tilde{\omega}_j^2,1/m_j),\; j=1,2.
\end{eqnarray}
We consider the fundamental commutation relations  in NC-space  as \cite{bastos1}
\begin{equation}\label{NCspacecommutation}
	[	\hat{\tilde{X}}_{\alpha}, \hat{\tilde{X}}_{\beta}]= i\hbar_e \tilde{J}_{\alpha\beta}= - (\tilde{\Sigma}_y)_{\alpha\beta},
\end{equation}
where the deformed symplectic matrix $\tilde{J}$ and the effective Planck constant $\hbar_e$ reads
\begin{eqnarray}
	\tilde{J} = \left(\begin{array}{cc}
		J_2 &  \frac{1}{\hbar_e}\Pi_{\theta\eta}\\
		-\frac{1}{\hbar_e}\Pi_{\theta\eta} &  J_2
	\end{array}\right),\; \mbox{with}\; 
	\Pi_{\theta\eta}= \left(\begin{array}{cc}
		\theta &  0\\
		0 & \eta
	\end{array}\right),\; 
	\hbar_{e}=\hbar(1+\frac{\theta\eta}{4\hbar^2}).
\end{eqnarray}
Here $\theta$ and $\eta$ are the position position NC-parameter and momentum-momentum NC-parameter, respectively. $\hbar$ is the Planck constant, and $J_2$ is the usual symplectic matrix, which encodes the commutation relations in the usual commutative space 	$\hat{X}=(\hat{X}_1, \hat{X}_2,\hat{X}_3,\hat{X}_4)^T=(\hat{x}_1, \hat{p}_1,\hat{x}_2,\hat{p}_2)^T$ as follows.
\begin{eqnarray}\label{Cspacecommutation}
	[	\hat{X}_{\alpha}, \hat{X}_{\beta}]=i\hbar J_{\alpha\beta}= -\hbar(\Sigma_y)_{\alpha\beta}, \;
	\mbox{with}\;
	J=\mbox{diag}(J_2,J_2),\\ \Sigma_j=\mbox{diag}(\sigma_j, \sigma_j),\; \mbox{for}\; j=x,y,z.\nonumber
\end{eqnarray}
Here the symplectic matrix $J_2$ and Pauli matrices are represented by
\begin{eqnarray}
	\sigma_x =\left(\begin{array}{cc}
		0 & 1\\
		1& 0
	\end{array}\right),\;
	\sigma_y =\left(\begin{array}{cc}
		0 & -i\\
		i& 0
	\end{array}\right),\;
	\sigma_z =\left(\begin{array}{cc}
		1 & 0\\
		0& -1
	\end{array}\right),\;
	J_2 = \left(\begin{array}{cc}
		0 & 1\\
		-1& 0
	\end{array}\right).
\end{eqnarray}

The NC-space co-ordinates ($\hat{\tilde{X}}$) are connected to the  commutative space co-ordinates ($\hat{X}$) through the Darboux transformation ($\Upsilon_D$) given by the Bopp's shift \cite{NCent1}
\begin{eqnarray}\label{cncconnection}
	\hat{\tilde{X}}=\Upsilon_D \hat{X},\;\;
	\mbox{with}\;
	\Upsilon_D =\left(\begin{array}{cc}
		\mathbb{I}_2 & -\frac{1}{2\hbar}\Pi_{\theta\eta}J_2 \\
		\frac{1}{2\hbar}\Pi_{\theta\eta}J_2 & \mathbb{I}_2 
	\end{array}\right).
\end{eqnarray}
The notation $\mathbb{I}_n$ stands for $n\times n$ identity matrix.
Here our concern is to have a valid Darboux transformation. Thus we restrict the parameter choice such that $\frac{\theta\eta}{4\hbar^2} <1$. In other words, we restrict the determinant of $\Upsilon_D $ is nonzero ($\Delta_{\Upsilon_D}\neq 0$); i.e., $\Upsilon_D \in GL(4, R)$.
$J$ is connected with the deformed symplectic matrix $\tilde{J}$ through 
\begin{equation}\label{JncJcconnection}
	\hbar_e \tilde{J} = \hbar \Upsilon_D J\Upsilon_D^T .
\end{equation}
Since the quantum mechanical formalism are well established in commutative space, it is customary to convert the NC-space system in the usual commutative space system through ~\eqref{cncconnection} for computational purpose.
Using ~\eqref{cncconnection}, one can see that the NC-space Hamiltonian ~\eqref{NCShamiltonian} is equivalent to the usual commutative space Hamiltonian 
\begin{eqnarray}\label{Hamiltonianmatrix}
	\hat{H}=\frac{1}{2} \hat{X}^T \mathcal{H}\hat{X},\; \mbox{with}\; \mathcal{H}=\Upsilon_D^T \mathcal{H}_{\mathrm{nc}}\Upsilon_D.
\end{eqnarray}
Explicitly written
\begin{eqnarray}
	\mathcal{H}=\left(\begin{array}{cc}
\mathcal{H}_1 & \mathcal{H}_{12}\\
\mathcal{H}_{12}^T & \mathcal{H}_2
	\end{array}\right),\;
\mbox{with}\; \mathcal{H}_{12}=\left(\begin{array}{cc}
	0 & -2\nu_2 \\
	2\nu_1 & 0
\end{array}\right),\; \mathcal{H}_j=\mbox{Diag}(\mu_j\omega_j^2,1/\mu_j); j=1,2
.
\end{eqnarray}
Explicit forms of the  parameters are given by
\begin{eqnarray}
	\frac{1}{\mu_1} &=& \frac{1}{m_1}+ \frac{1}{4\hbar^2} m_2\tilde{\omega}_2^2\theta^2, \;
	\frac{1}{\mu_2} = \frac{1}{m_2}+ \frac{1}{4\hbar^2} m_1\tilde{\omega}_1^2\theta^2, \label{para1}\\
	\alpha_1&=& \mu_1\omega_1^2=m_1\tilde{\omega}_1^2 + \frac{\eta^2}{4\hbar^2 m_2} ,\;
	\alpha_2  = \mu_2\omega_2^2=m_2\tilde{\omega}_2^2 + \frac{\eta^2}{4\hbar^2 m_1}, \label{para2}\\
	\nu_1 & =& \frac{1}{4m_1\hbar} (\eta + m_1 m_2\tilde{\omega}_2^2 \theta),\;
	\nu_2 = \frac{1}{4m_2\hbar} (\eta + m_1 m_2\tilde{\omega}_1^2 \theta). \label{para3}
\end{eqnarray}
Note that for isotropic oscillator $\mu_1=\mu_2,\; \alpha_1=\alpha_2,\; \nu_1=\nu_2$. For an isotropic oscillator, one can identify the off-diagonal terms constitute the angular momentum operator. \\
One can diagonalize the bilinear Hamiltonian ~\eqref{Hamiltonianmatrix} with a normal coordinate system, keeping the symplectic structure $Sp(4,\mathbb{R})$ intact. First we note that, the symplectic eigenvalues \cite{williamson1,williamson2} of $\mathcal{H}$ are the ordinary eigenvalues of $
	\mathcal{H}_J=J\mathcal{H}$.
Since $	\mathcal{H}_J$ is not symmetric, the left and right eigenvectors of $	\mathcal{H}_J$ are not the same. However, left and right eigenvalues are identical. The characteristic polynomial $	P_{\mathcal{H}_J}(\lambda)$ of $\mathcal{H}_J$ has four distinct purely imaginary roots for the  real parameters $\mu_j,\omega_j,\nu_j$:
\begin{equation}\label{lambdaformexplicit}
	\lambda \in \{\mp i\lambda_j, j=1,2\vert \lambda_1=\sqrt{(\Delta-D)/2}, \lambda_2=\sqrt{(\Delta+D)/2} \},
\end{equation}
where $\Delta=\Delta_{\mathcal{H}_1}+\Delta_{\mathcal{H}_2}+2\Delta_{\mathcal{H}_{12}}$, and the discriminant $D=\sqrt{\Delta^2-4\Delta_{\mathcal{H}_J}}$ is given by
\begin{eqnarray}\label{discriminant}
	D^2= (\omega_1^2-\omega_2^2)^2 + 16\nu_1\nu_2 (\omega_1-\omega_2)^2 + 16 \left(\sqrt{\frac{\mu_1}{\mu_2}}\omega_1\nu_1 + \sqrt{\frac{\mu_2}{\mu_1}}\omega_2\nu_2\right)^2\ge 0.
\end{eqnarray}
Here we have used the notation $\Delta_A=\mbox{Det}(A)$.
Note that $D=0$ only for parameter values $\omega_1=\omega_2,\; \nu_1=\nu_2=0$, which corresponds to the isotropic oscillator in commutative space. We shall consider $D>0$ for our present study.

If $\chi_{lj}$ is the left eigenvector corresponding to the eigenvalue $-i\lambda_j$ of $\mathcal{H}_J$, i.e., $\chi_{lj}\mathcal{H}_J=-i\lambda_j \chi_{lj}$, then the direct computation gives
\begin{eqnarray}
	\chi_{lj}=k_{j}(i \kappa_{j,1}, \kappa_{j,2}, \kappa_{j,3}, i\kappa_{j,4});\; j=1,2.
\end{eqnarray}
Here $k_{j}$ is the normalization constant, and the real parameters $\kappa_{j,\alpha}^{r} $ and $\kappa_{j,\alpha}$ reads
\begin{eqnarray}
	\kappa_{j,1} &=& -2\mu_1\lambda_j (\mu_1\nu_1\omega_1^2 + \mu_2\nu_2\omega_2^2),\;
	\kappa_{j,2} = 2(\mu_2\nu_2\omega_2^2 - 4\mu_1\nu_1^2 \nu_2 + \mu_1\nu_1\lambda_j^2), \label{kappaj12}\\
	\kappa_{j,3} &=& \mu_1(4\mu_1\nu_1^2\omega_1^2 - \mu_2 \omega_1^2 \omega_2^2 + \mu_2\omega_2^2 \lambda_j^2),\;
	\kappa_{j,4} = -\mu_1\lambda_j (\omega_1^2 + 4\nu_1\nu_2 -\lambda_j^2). \label{kappaj34}
\end{eqnarray}
The left eigenvector corresponding to the eigenvalue $i\lambda_j$ is given by $	\chi_{lj}^*$. The right eigenvector $\chi_{rj}$ corresponding to the eigenvalue $-i\lambda_j$   may be obtained through   $\chi_{rj}=-\Sigma_y \chi_{lj}^\dagger$. Normalization condition $\chi_{lj}\chi_{rj}=1$ yields
\begin{equation}\label{normalizationofchi}
	\vert k_{j}\vert= 1/\sqrt{2(\kappa_{j,3}\kappa_{j,4}- \kappa_{j,1} \kappa_{j,2})};\; j=1,2. 
\end{equation}
Similarity transformation which diagonalizes $\mathcal{H}_J$, i.e., 
\begin{equation}
	\mathcal{H}_{JD}=\mbox{diag}(-i\lambda_1,i\lambda_1,-i\lambda_2,-i\lambda_2)= Q^{-1}\mathcal{H}_J Q,
\end{equation}
is given by the following matrices
\begin{equation}
	Q=(\chi_{r1},\chi_{r1}^*,\chi_{r2},\chi_{r2}^*),\;	Q^{-1}=(\chi_{l1}^T,\chi_{l1}^{*T},\chi_{l2}^T,\chi_{l2}^{*T})^T.
\end{equation}
The diagonal representation of $\mathcal{H}_J$ enables to  define the normal co-ordinates through
\begin{equation}\label{a1a2}
	\hat{A}=(\hat{a}_1,\hat{a}_1^\dagger,\hat{a}_2,\hat{a}_2^\dagger)^T=\frac{1}{\sqrt{\hbar}}Q^{-1}\hat{X}.
\end{equation}
Using the normalization condition ~\eqref{normalizationofchi}, it follows 
\begin{equation}\label{algebraaadagger}
	[\hat{a}_1,\hat{a}_1^\dagger]=	[\hat{a}_2,\hat{a}_2^\dagger]=1.
\end{equation}
On the other hand, the orthogonality condition $\chi_{l1}\chi_{r2}=\chi_{l2}\chi_{r1}=0$ is equivalent to
\begin{equation}\label{algebraa1a2}
	[\hat{a}_1,\hat{a}_2]=0.
\end{equation}
The algebra ~\eqref{algebraaadagger} and ~\eqref{algebraa1a2} confirm that $\hat{a}_j$ and $\hat{a}_j^\dagger$ are annihilation and creation operators.
In term of the normal co-ordinates the Hamiltonian ~\eqref{Hamiltonianmatrix} reads
\begin{equation}\label{HM}
	\hat{H}_M= \sum_{j=1}^{2}\hbar\lambda_j(\hat{n}_j+1/2),\;\mbox{with}\; \hat{n}_j=\hat{a}_j^\dagger \hat{a}_j. 
\end{equation}
The ground state of ~\eqref{HM} satisfy the property \cite{NCent3}
\begin{equation}\label{groundstatedefnforanisotropic}
	\hat{a}_1\vert 0,0\rangle =\hat{a}_2\vert 0,0\rangle =0.
\end{equation}
In position representation ($\psi_{0,0}(x_1,x_2)=\langle x_1,x_2\vert 0,0\rangle$), the equation ~\eqref{groundstatedefnforanisotropic} reads
\begin{equation}\label{groundstateforanisotropic}
	(U_xx-i\hbar U_p\partial_{x})\psi_{0,0}(x_1,x_2)=0,
\end{equation}
where
\begin{eqnarray}
	x=\left(\begin{array}{c}
		x_1 \\
		x_2
	\end{array}\right),\; 
	\partial_x=\left(\begin{array}{c}
		\frac{\partial}{\partial x_1} \\
		\frac{\partial}{\partial x_2}
	\end{array}\right),\;
	U_x=\left(\begin{array}{cc}
		i\kappa_{11} & \kappa_{31}\\
		i\kappa_{12} & \kappa_{32}
	\end{array}\right),\;
	U_p=\left(\begin{array}{cc}
		\kappa_{21} & i\kappa_{41}\\
		\kappa_{22} & i\kappa_{42}
	\end{array}\right).
\end{eqnarray}
We take the following ansatz for the solution of ~\eqref{groundstateforanisotropic}.
\begin{equation}\label{ansatzpsi00anisotropic}
	\psi_{0,0}(x_1,x_2)= \mathcal{N}_0 e^{-\frac{1}{2}x^T\Lambda_{\mathrm{ancs}}x}.
\end{equation}
Here, $\mathcal{N}_0$ is the normalization constact, and
\begin{eqnarray}
	\Lambda_{\mathrm{ancs}}=\left(\begin{array}{cc}
		\Lambda_{11} & \Lambda_{12}\\
		\Lambda_{12} & \Lambda_{22}
	\end{array}\right).
\end{eqnarray}
Using ~\eqref{ansatzpsi00anisotropic} in ~\eqref{groundstateforanisotropic} we get
\begin{equation}
	\Lambda_{\mathrm{ancs}}=\frac{i}{\hbar} U_p^{-1}U_x.
\end{equation}
Explicitly written,
\begin{eqnarray}
	\Lambda_{11} &=& \frac{\kappa_{41}\kappa_{12}-\kappa_{42}\kappa_{11}}{\hbar(\kappa_{21}\kappa_{42}-\kappa_{22}\kappa_{41})} =\Lambda_{11r},\;\;
	\Lambda_{22} = \frac{\kappa_{21}\kappa_{32}-\kappa_{22}\kappa_{31}}{\hbar(\kappa_{21}\kappa_{42}-\kappa_{22}\kappa_{41})}=\Lambda_{22r}, \label{Lambda11form}\\
	\Lambda_{12} &=& \frac{i(\kappa_{42}\kappa_{31}-\kappa_{41}\kappa_{32})}{\hbar(\kappa_{21}\kappa_{42}-\kappa_{22}\kappa_{41})}=\frac{i(\kappa_{12}\kappa_{21}-\kappa_{22}\kappa_{11})}{\hbar(\kappa_{21}\kappa_{42}-\kappa_{22}\kappa_{41})}\label{Lambda12form}=i\Lambda_{12c}.
\end{eqnarray}
In particular, $\Lambda_{11}$ and $\Lambda_{22}$ are real, whereas $\Lambda_{12}$ is purely imaginary. 
\subsection{Noise matrix and separability criterion}
Wigner distribution ($W$) and the density operator $(\hat{\rho})$ are related
through the definition \cite{weedbrook}
\begin{equation}
	W(X_c) =\frac{1}{\pi^2\hbar^2}\int d^2 x' \langle x-x'\vert \hat{\rho} \vert x+x'\rangle e^{2ix'.p/\hbar},
\end{equation}
where $X_c=(X_1,X_2,X_3,X_4)=(x_1,p_1,x_2,p_2)^T$ is the classical co-ordinate  vector. We define 
\begin{eqnarray}
	\Delta \hat{X}_\alpha &=& \hat{X}_\alpha -\langle \hat{X}_\alpha\rangle_\rho,  \mbox{with}\; \langle \hat{X}_\alpha\rangle_\rho = \mbox{Tr}(\hat{X}_\alpha\hat{\rho}); \;\alpha=1,2,3,4.\\
	\Delta X_\alpha &=& X_\alpha -\langle X_\alpha\rangle_W,\; \mbox{with}\; \langle X_\alpha\rangle_W = \int   W(X_c)X_\alpha d^4X_c .
\end{eqnarray}
The four components $\Delta \hat{X}$ satisfy the same commutation relations as $\hat{X}$. Moreover, the phase-space average $\langle X_\alpha\rangle_W$ with respect to the Wigner distribution $W$ is equal to the average $\langle \hat{X}_\alpha\rangle_\rho$ with respect to the density operator $\hat{\rho}$. We define the covariance matrix $\mathcal{V}_c$ through the matrix elements
\begin{equation}\label{covdefn}
	\mathcal{V}_{\alpha\beta}=\frac{1}{2} \langle \{\Delta \hat{X}_\alpha, \Delta \hat{X}_\beta\}\rangle =\mbox{Tr}(\frac{1}{2}\{\Delta \hat{X}_\alpha, \Delta \hat{X}_\beta\}\hat{\rho})=\int \Delta X_\alpha \Delta X_\beta W(X_c)d^4X_c.
\end{equation}
Since $\mathcal{V}_{\mathrm{c}}$ is symmetric, we can  write it in the block form
\begin{eqnarray}\label{cspacecovariancematrix}
	\mathcal{V}_{\mathrm{c}}=\left(\begin{array}{cc}
		V_{11} & V_{12}\\
		V_{12}^T & V_{22}
	\end{array}\right).
\end{eqnarray}
Using the connection ~\eqref{cncconnection} between the commutative space and NC-space co-ordinates, one can see that the NC-space covariance matrix $\mathcal{V}_{nc}$ is given by
\begin{equation}\label{VncVcconnection}
	\tilde{\mathcal{V}}_{\mathrm{nc}}= \Upsilon_D  \mathcal{V}_{\mathrm{c}}\Upsilon_D^T .
\end{equation}
The fundamental commutation relations ~\eqref{Cspacecommutation} of commutative space implies that a bonafide covariance matrix must satisfy the Robertson-Schr\"{o}dinger uncertainty principle (RSUP)\cite{weedbrook}
\begin{equation}\label{RSUPc}
	\mathcal{V}_{\mathrm{c}}+\frac{i}{2}\hbar J \ge 0 .
\end{equation}
In NC-space, the equivalent statement for the RSUP reads \cite{Moyal2,ncssim1,Gouba1,NCent1}
\begin{equation}\label{RSUPnc}
	\tilde{\mathcal{V}}_{\mathrm{nc}} + \frac{i}{2} \hbar_e \tilde{J} \ge 0.
\end{equation}
A generic local transformation $S_1 \bigoplus S_2$, acts on $\mathcal{V}_{\mathrm{c}}$ as
\begin{equation}
	V_{jj}\to S_j V_{jj} S_j^T ,\; 	V_{12}\to S_1 V_{12}S_2^T;\; \mbox{with}\; S_j \in Sp(2,\mathbb{R}),\;  j=1,2.
\end{equation}
One can identify that following four quantities are local  invariant with respect to transformation belonging to the $Sp(2,\mathbb{R})\bigotimes Sp(2,\mathbb{R}) \subset Sp(4,\mathbb{R})$.
\begin{eqnarray}
	\Delta_j=\mbox{Det}(V_{jj}),	\Delta_{12}=\mbox{Det}(V_{12}),  \Delta_{\mathcal{V}_{\mathrm{c}}} = \mbox{Det}(\mathcal{V}_{\mathrm{c}}),
	\tau_{v}=\mbox{Tr}(V_{11} J_2 V_{12} J_2 V_{22} J_2 V_{12}^T J_2).
\end{eqnarray}
Using Williamson's theorem \cite{williamson1,williamson2}, one can show that the RSUP ~\eqref{RSUPc} can be rewritten as $Sp(2,\mathbb{R})\bigotimes Sp(2,\mathbb{R})$ invariant statement \cite{simon1,simon2,simon3}
\begin{equation}\label{covariancersupc}
	\Delta_1 \Delta_2 + (\hbar^2/4- \Delta_{12})^2 -\tau_{v}\ge \hbar^2( \Delta_1+\Delta_2)/4.
\end{equation}
Under mirror reflection (Peres-Horodecki partial transpose) $\Delta_1$,  $\Delta_2$ and $\tau_v$ remain invariant; whereas $\Delta_{12}$ flips sign. Therefore, the requirement that the covariance matrix of a separable state has to obey the following necessary condition \cite{simon1,simon2,simon3}.
\begin{equation}\label{separabilityc}
	Ps=	\Delta_1 \Delta_2 + (\hbar^2/4-\vert \Delta_{12}\vert)^2 - \tau_{v}- \hbar^2( \Delta_1+\Delta_2)/4 \ge 0,
\end{equation}
which turns out to be sufficient for all bipartite-Gaussian state.\\
Now, for a generic ground state ~\eqref{groundstatedefnforanisotropic}, which has the form in position representation as
$\psi(x_1,x_2)=\mathcal{N}_0e^{-\frac{1}{2}x^T\Lambda x}$, corresponds to 
the Wigner distribution
\begin{equation}\label{WX}
	W(X)= \frac{1}{\pi^2\hbar^2} \exp\{ - x^T (\Lambda_r + \Lambda_c \Lambda_r^{-1}\Lambda_c^T)x - \frac{1}{\hbar^2} p^T\Lambda_r^{-1}p  -\frac{1}{\hbar} (x^T \Lambda_c \Lambda_r^{-1} p + p^T \Lambda_r^{-1}\Lambda_c^T x) \}.
\end{equation}
Here the following notation for the matrix $\Lambda$ and its elements $\Lambda_{jk}$ have been used.
\begin{eqnarray}
	\Lambda=[\Lambda_{jk}]_{j,k=1}^{2}= \Lambda_r +i\Lambda_c= [\Lambda_{jkr}]_{j,k=1}^{2}+i[\Lambda_{jkc}]_{j,k=1}^{2},
\end{eqnarray}
where $\Lambda_{jkr}=\Re(\Lambda_{jk}),\Lambda_{jkc}=\Im(\Lambda_{jk})$, and $x=(x_1,x_2)^T,\; p=(p_1,p_2)^T$.
The expectation values through the Wigner distribution $W(X)$, provide the covariance matrix $\mathcal{V}_{\mathrm{c}}$ as
\begin{eqnarray}\label{vc}
	\mathcal{V}_{\mathrm{c}}=  \frac{\hbar}{2}\left(\begin{array}{cc}
		\sigma_{11} & \sigma_{12}\\
		\sigma_{12}^T & \sigma_{22}
	\end{array}\right),
\end{eqnarray}
with
\begin{eqnarray}\label{formofsigmaij}
	\sigma_{11}=\left(\begin{array}{cc}
		\frac{1}{\hbar\Lambda_{11r}} & 0\\
		0 & \frac{\hbar\Delta_\Lambda}{\Lambda_{22r}}
	\end{array}\right),\;
	\sigma_{22}=\left(\begin{array}{cc}
		\frac{1}{\hbar\Lambda_{22r}} & 0\\
		0 & \frac{\hbar\Delta_\Lambda}{\Lambda_{11r}}
	\end{array}\right),\;
	\sigma_{12}=\left(\begin{array}{cc}
		0& -	\frac{\Lambda_{12c}}{\Lambda_{11r}} \\
		-\frac{\Lambda_{12c}}{\Lambda_{22r}} &0
	\end{array}\right),
\end{eqnarray}
where $\Delta_\Lambda= \Lambda_{11r}\Lambda_{22r}+\Lambda_{12c}^2$.
Using  ~\eqref{formofsigmaij} in the generalized Peres-Horodecki separability criterion ~\eqref{separabilityc} we get the following constraint on the parameters.
\begin{equation}
	-\Lambda_{11r} \Lambda_{22r} \Lambda_{12c}^2 \ge \Lambda_{12c}^2 \Lambda_{11r}\Lambda_{22r}\implies
	\Lambda_{11r}\Lambda_{22r}\Lambda_{12c}=0.
\end{equation}
Since $\Lambda_{11r}$ and $\Lambda_{22r}$ are nonzero, the separability of states implies $\Lambda_{12c}=0$, i.e.,
\begin{equation}\label{constrainteqn}
	(\lambda_2-\lambda_1)(\mu_2\nu_2\omega_2^2 -4\mu_1\nu_1^2\nu_2-\mu_1\nu_1\lambda_1\lambda_2)=0.
\end{equation}
However, according to ~\eqref{discriminant} and ~\eqref{lambdaformexplicit}, we consider $\lambda_1\neq \lambda_2$. Therefore, ~\eqref{constrainteqn} holds for
\begin{equation}
	(\mu_1^2\omega_1^2\nu_1^2-\mu_2^2\omega_2^2\nu_2^2)(\omega_2^2-4\mu_1\nu_1^2/\mu_2)=0.
\end{equation}
From the physical viewpoint, we choose $\theta,\eta \le \hbar$. In other words, $\theta\eta \neq 4\hbar^2$, which means $\omega_2^2 \neq 4\mu_1\nu_1^2/\mu_2$. Therefore, the only possibility for the separable states is satisfied by the constraint $\mu_1\nu_1\omega_1 =\mu_2\nu_2\omega_2$, which is equivalent to the following equation in terms of original parameters.
\begin{eqnarray}\label{sep1}
	(4\hbar^2/m_{12}+\tilde{\omega}_1^2\theta^2) (\eta/m_{12}+ \tilde{\omega}_2^2\theta)^2 (\eta^2/m_{12}+ 4\hbar^2\tilde{\omega}_1^2) 
	=  (4\hbar^2/m_{12}+\tilde{\omega}_2^2\theta^2) \nonumber \\
	(\eta/m_{12}+ \tilde{\omega}_1^2\theta)^2 (\eta^2/m_{12}+ 4\hbar^2\tilde{\omega}_2^2) ,\; \mbox{with}\; m_{12}=m_1.m_2.
\end{eqnarray}
The condition ~\eqref{sep1} is trivially satisfied for commutative space ($\theta,\eta\to 0$), and as well as for isotropic oscillator ($\tilde{\omega}_1=\tilde{\omega}_2$) in noncommutative (NC) space. Therefore, the entanglement between the coordinate degrees of freedom is not the sole property of the noncommutativity of space. It depends on both the NC-parameters ($\theta,\eta$) and the anisotropy of oscillator frequency ($\tilde{\omega}_1\neq\tilde{\omega}_2$). If we recall that the off-diagonal term for the isotropic oscillator in NC-deformed space under consideration is merely an angular momentum operator, which commutes with the rest of the Hamiltonian in the equivalent commutative space. Thus, the ground state (Gaussian) is just a product of two ordinary oscillators, resulting in separable states. However, for an anisotropic oscillator, the deformed angular momentum operator does not commute with the rest portion of the Hamiltonian, which results in the entangled state. Thus, we can conclude that the entanglement is created via the NC-deformation of space for an anisotropic oscillator.   Moreover, even in NC-space, an anisotropic oscillator also supports separable states. For instance, let us consider, without loss of generality $m_1m_2=1, \hbar=1, \mu=1,\nu=1$, which provides the relation $\tilde{\omega}_2=1/\tilde{\omega}_1$ between the oscillator frequencies for which it supports separable states.
All the other frequencies admit entanglement between coordinate degrees of freedom. Since the separability of the bipartite Gaussian state for an anisotropic oscillator in NCS is satisfied for only a very special choice of anisotropic parameters, it is worth mentioning that the bipartite Gaussian state is almost always entangled in NC-space.
\section{Experimental scheme: optomechanical scheme and Stokes-like measurements}
Until now, in the present paper, we have discussed the entanglement generated dynamically. In particular, we devised the entanglement through Bopp's shift. In a real-life scenario, however, we are  ignorant about whether the background space is commutative or noncommutative. We have to perform an experiment to determine whether the background phase-space is deformed.  In this section, we discuss a gedankenexperiment through an optomechanical scheme along with a Stokes-like measurement. The experimental results enable us to write down moments of operators just by measuring the ensemble average of photocurrents in photodetectors. Additionally, this provides an experimental scheme to determine NC-parameters in terms of measured photocurrents.
\begin{figure}
	\includegraphics[width=8cm]{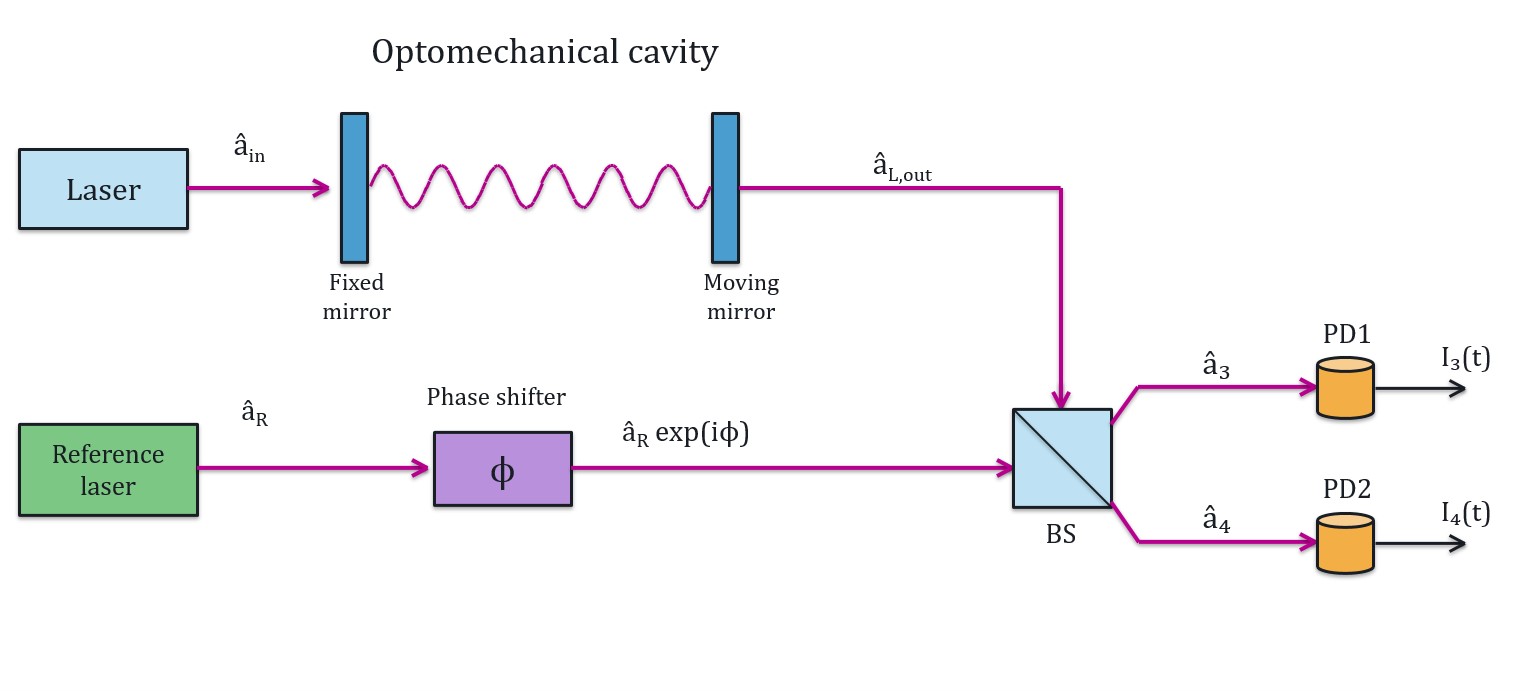}
	\caption{{\bf Optomechanical interaction and Stokes-like measurement.} An input optical beam $\hat{a}_L=\hat{a}_{in}$ enters the cavity and interacts with the mechanical oscillator (arm of the mirror). Output beam mixed with phase-shifted reference beam. Produced photocurrents (proportional to photon numbers) are measured through photodetectors.   }
	\label{opto}
\end{figure}
Suppose an input optical mode $\hat{a}_{L}$, with frequency $\omega_L$, and having photon number operator $\hat{n}_{L}=\hat{a}_{L}^\dagger \hat{a}_{L}$, enters the optical cavity and interacts with the mechanical oscillator $\hat{H}_M$ of ~\eqref{HM}. $\hat{a}_L$ interacts dispersively with the collective NC-mechanical position \cite{expt2,lightmatter3,cavity1}
\begin{equation}\label{collectivenccoordinate}
	\hat{\mathcal{X}} = g_1\hat{\tilde{x}}_1 + g_2\hat{\tilde{x}}_2 =\sum_{j=1}^2 (f_j \hat{a}_j + f_j^*\hat{a}_j^\dagger),
\end{equation}
where $g_j (j=1,2)$ are calibrated optical coupling strengths. The mechanical normal modes $\hat{a}_j$ $(j=1,2)$, are given by ~\eqref{a1a2}, and the value of $f_j$ are given by
\begin{equation}
f_j= \sqrt{\hbar}\kappa_j [g_2(\kappa_{j,4}-\epsilon \kappa_{j,1}) +i g_1( \kappa_{j,2} +  \epsilon  \kappa_{j,3} )],\; \mbox{with}\; \epsilon=\theta/(2\hbar),\; j=1,2.
\end{equation}
$\kappa_{j,l}$ are given in ~\eqref{kappaj12} and ~\eqref{kappaj34}. Following the standard radiation-pressure interaction employed in cavity optomechanics \cite{lightmatter3,cavity1}, we consider a dispersive coupling between $\hat{n}_L$ and the collective displacement of the two mechanical normal modes. If we assume a constant coupling $g$ for $\hat{\mathcal{X}}$ and $\hat{n}_L$, during an optical short-pulse interaction of duration $\tau$, the total Hamiltonian reads \cite{expt2,lightmatter3,cavity1}
\begin{equation}\label{optomechint}
\hat{H}_{\mathrm{tot}}=\hat{H}_0+\hat{H}_{\mathrm{int}},\; \mbox{with}\; \hat{H}_0=\hat{H}_M+\hbar\omega_L\hat{n}_L,\; \hat{H}_{\mathrm{int}}=-\hbar g \hat{n}_{L}\hat{\mathcal{X}}.
\end{equation}
 The total Hamiltonian  $\hat{H}_{tot}$ commutes with $\hat{n}_{L}$. Hence, photon number is conserved. In other words, we consider an ideal lossless optomechanical interaction. With respect to the free Hamiltonian $\hat{H}_0$, the interaction Hamiltonian becomes
 \begin{equation}
 	\hat{H}_{I}(t)=-\hbar g \hat{n}_L\sum_{j=1}^{2}(f_je^{-i\lambda_j t}\hat{a}_j+ f_j^*e^{i\lambda_j t}\hat{a}_j^\dagger).
 \end{equation}
The Hamiltonians at two different times do not commute. In particular,
\begin{equation}\label{HITcommutator}
	[\hat{H}_{I}(t_1),\hat{H}_{I}(t_2)]=2if_{g\lambda}(t_1,t_2)\hat{n}_{L}^2,\; \mbox{with}\; 
	f_{g\lambda}(t_1,t_2)=\hbar^2g^2\sum_{j=1}^{2}\vert f_j\vert^2 \sin(\lambda_j(t_2-t_1)).
\end{equation}
Since ~\eqref{HITcommutator} is proportional to $\hat{n}_L^2$,  the time-evolution operator $\hat{U}_{I}(t)$, which is given by the time-ordering operator $\mathcal{T}$ as \cite{entanglement3}
\begin{equation}
	\hat{U}_{I}(t)=\mathcal{T}e^{-i\hbar^{-1}\int_{0}^{t}\hat{H}_{I}(s)ds},
\end{equation}
can be evaluated analytically in a straightforward manner. In particular,
\begin{equation}
	\hat{U}_{I}(t)=e^{i\Phi_K(t)\hat{n}_L^2}\prod_{j=1}^{2}\mathcal{D}_j(\hat{n}_L\zeta_j(t)),
\end{equation}
where the displacement operator is defined as
\begin{equation}
	\mathcal{D}_j(\zeta)=e^{\zeta\hat{a}_j^\dagger -\zeta^* \hat{a}_j}.
\end{equation}
$\zeta_j$ and the effective Kerr-nonlinear optical geometric phase $\Phi_K(t)$ are given by \cite{entanglement3,expt2,expt3}
\begin{equation}\label{Kerrphase}
	\zeta_j=\frac{gf_j^*}{\lambda_j}(e^{i\lambda_j t}-1),\; 
	\Phi_K(t)=\sum_{j=1}^{2}\frac{g^2\vert f_j\vert^2}{\lambda_j^2}(\lambda_jt-\sin(\lambda_j t)).
\end{equation}
The mechanical ground state displacement overlap is \cite{entanglement3}
\begin{equation}\label{AMexpt}
	\langle 0,0\vert \mathcal{D}_1(\zeta_1)\mathcal{D}_2(\zeta_2)\vert 0,0\rangle =e^{-A_{M}(t)},\;
	\mbox{with}\; A_M(t)=\frac{1}{2}\sum_{j=1}^{2}\vert \zeta_j\vert^2.
\end{equation}
After the optomechanical interaction, the output mode is 
\begin{equation}
	\hat{a}_S=\hat{a}_{L,\mathrm{out}}(t)=\hat{U}_I^\dagger (t) \hat{a}_L \hat{U}_I(t)= 
	e^{-i\omega_L t}e^{i\Phi_K(t)(2\hat{n}_L+1)}\hat{a}_L \prod_{j=1}^2 \mathcal{D}_j(\zeta_j(t)).
\end{equation}
 If we maintain the experimental set up at a sufficient low temperature (below $4$ Kelvin), then the oscillator state may be well approximated by the ground state, in particular, a coherent state ~\eqref{ansatzpsi00anisotropic}. If we take the input state to be a coherent state $\vert \alpha_l\rangle_L$, then the initial density operator is given by
\begin{equation}
	\hat{\rho}_{\mathrm{tot}}(0) =\vert \alpha_l\rangle_L \langle\alpha_l\vert \otimes \vert 0,0\rangle_M \langle 0,0\vert.
\end{equation}
The optical moments of the output mode $\hat{a}_S$ are given by
\begin{equation}\label{M1M2theory}
	\mathcal{M}_1(t)=\langle \hat{a}_S\rangle = R_{M_1}(t)e^{i\delta_1(t)},\;
		\mathcal{M}_2(t)=\langle \hat{a}_S^2\rangle = R_{M_2}(t)e^{i\delta_2(t)},
\end{equation}
where
\begin{eqnarray}
R_{M_1}(t) &=& \vert \alpha_l\vert e^{-A_M+ \vert\alpha_l\vert^2(\cos2\Phi_K -1)},\;
\delta_1(t)= \arg\alpha_l +\Phi_K +\vert\alpha_l\vert^2\sin2\Phi_K, \label{RM1}\\
R_{M_2}(t) &=& \vert \alpha_l\vert^2 e^{-4A_M+ \vert\alpha_l\vert^2(\cos4\Phi_K -1)},\;
\delta_2(t)= 2\arg\alpha_l +4\Phi_K +\vert\alpha_l\vert^2\sin4\Phi_K. \label{RM2}
\end{eqnarray}
Note that, since the photon number ($n_L = \langle \hat{a}_S^\dagger\hat{a}_S\rangle = \vert\alpha_l\vert^2$) is conserved, the optical covariance matrix for the output beam $\hat{a}_S$ can be determined from $\mathcal{M}_1$, $\mathcal{M}_2$ and $n_L$ \cite{entanglement3,stokes1,weedbrook}. In particular, if we define optical quadrature operators
\begin{equation}
	\hat{x}_S=(\hat{a}_S + \hat{a}_S^\dagger)/\sqrt{2}, \;  \hat{p}_S=(\hat{a}_S-\hat{a}_S^\dagger)/(i\sqrt{2}),
\end{equation}
the optical covariance matrix reads
\begin{equation}
	V_S=\left(\begin{array}{cc}
V_{S,11} & V_{S,12}\\
V_{S,12} & V_{S,22}
	\end{array}\right) = \left(\begin{array}{cc}
	\langle \hat{x}_S^2\rangle -\langle \hat{x}_S\rangle^2 & \langle \hat{x}_S\hat{p}_S\rangle -\langle \hat{x}_S\rangle\langle \hat{p}_S\rangle\\
	\langle \hat{x}_S\hat{p}_S\rangle -\langle \hat{x}_S\rangle\langle \hat{p}_S\rangle & \langle \hat{p}_S^2\rangle -\langle \hat{p}_S\rangle^2
\end{array}\right),
\end{equation}
where 
\begin{eqnarray}
\langle \hat{x}_S\rangle &=& \sqrt{2}\Re(\mathcal{M}_1),\;
\langle \hat{p}_S\rangle=\sqrt{2}\Im(\mathcal{M}_1),\;
\langle \hat{x}_S\hat{p}_S\rangle= \langle \hat{x}_S\hat{p}_S + \hat{p}_S\hat{x}_S\rangle/2=\Im(\mathcal{M}_2),\\
\langle \hat{x}_S^2\rangle &=& 1/2+ n_L+\sqrt{2}\Re(\mathcal{M}_2),\;
\langle \hat{p}_S^2\rangle=1/2+ n_L-\sqrt{2}\Re(\mathcal{M}_2).
\end{eqnarray}
The notation $\Re(z)$ and $\Im(z)$ denote real and imaginary part of $z$, respectively.
After optomechanical interaction, $\hat{a}_S$ enters in the Stokes-measurement apparatus.
A reference (known) optical mode $\hat{a}_R$, which is phase-shifted with angle $\phi$ through a phase-shifter, is also passed through a $50-50$ beam splitter. 
The density operator of the reference beam in the number operator states $\{\vert n\rangle\}$ is given by \cite{entanglement3,stokes3}
\begin{equation}
	\hat{\rho}_r = \sum_{n_r=0}^{\infty} \frac{\bar{n}_r^n}{(\bar{n}_r+1)^{n+1}}\vert n_r\rangle \langle n_r\vert,
\end{equation}
where $\bar{n}_r$ denotes the mean photon numbers. $\hat{\rho}_r$  transforms under the displacement ($\hat{\mathcal{D}}(\alpha_r)=e^{\alpha_r \hat{a}_r^\dagger -\alpha_r^* \hat{a}_r}$) and squeezing operators ($\hat{\mathcal{S}}(\xi_r)=e^{\xi_r^* \hat{a}_r^2 -\xi_r (\hat{a}_r^\dagger)^2}$) as
\begin{equation}
	\hat{\rho}_r \mapsto \hat{\rho}_r^{DS} = \hat{U}_r \hat{\rho}_r\hat{U}_r^\dagger ,\; \hat{U}_r=\hat{\mathcal{D}}(\alpha_r)\hat{\mathcal{S}}(\xi_r).
\end{equation}
Hence, the expectation value of an operator $\hat{\mathcal{O}}$ on the displaced squeezed state $\hat{\rho}_r^{\mathrm{DS}}$ is given by
\begin{equation}\label{referenceexpectation}
	\langle \hat{\mathcal{O}}\rangle = \mbox{Tr}(\hat{\rho}_r^{\mathrm{DS}} \hat{\mathcal{O}})= \mbox{Tr}(\hat{\rho}_r \hat{U}_r^\dagger \hat{\mathcal{O}}\hat{U}_r).
\end{equation}
We parametrize the displacement parameters $\alpha_r$ and  sqeezing parameters $\xi_r$ as
\begin{equation}
	\alpha_r=d_r e^{i\beta_r} ,\;\; \xi_r=\theta_r e^{i\gamma_r}.
\end{equation}
Using $\hat{x}_r=(\hat{a}_R +\hat{a}_R^\dagger)/\sqrt{2}$, $\hat{p}_r=i(\hat{a}_R -\hat{a}_R^\dagger)/\sqrt{2}$ in ~\eqref{referenceexpectation}, we get the expectation values of the quadratures for the reference beams as follows \cite{stokes3}.
\begin{eqnarray}
	\langle \hat{x}_r\rangle &=& \sqrt{2}d_r \cos \beta_r,\; \langle \hat{p}_r\rangle = \sqrt{2} d_r \sin\beta_r. \\
	\langle \hat{x}_r^2\rangle &=& (\bar{n}_r +1/2) (e^{2\theta_r} \sin^2(\gamma_r/2)+ e^{-2\theta_r} \cos^2(\gamma_r/2)) + 2d_r^2 \cos^2\beta_r \\
	\langle \hat{p}_r^2\rangle &=& (\bar{n}_r +1/2) (e^{2\theta_r} \cos^2(\gamma_r/2)+ e^{-2\theta_r} \sin^2(\gamma_r/2)) + 2d_r^2 \sin^2\beta_r,\\
	\langle \hat{x}_r\hat{p}_r \rangle &=& d_r^2 \sin 2\beta_r - (\bar{n}_r +1/2)\sinh 2\theta_r \sin \gamma_r + i/2,\\
	\langle \hat{p}_r\hat{x}_r \rangle &=& d_r^2 \sin 2\beta_r - (\bar{n}_r +1/2)\sinh 2\theta_r \sin \gamma_r - i/2.
\end{eqnarray}
For simplicity, if we consider the reference beam with $\beta_r=\gamma_r=0$, i.e., $	\langle \hat{x}_r\rangle=\sqrt{2}d_r$, and $	\langle \hat{p}_r\rangle=0$, then the reference beam covariance matrix reads
\begin{equation}
	V_R=\mbox{Diag}(b_{R-}^2,b_{R+}^2),\; \mbox{with}\; b_{R\pm}^2 =(\bar{n}_r+1/2)e^{\pm 2\theta_r.}
\end{equation}
Following the standard quantum-optical treatment of balanced beam splitters and phase-sensitive interferometric detection \cite{stokes3,cavity2}, the signal and reference optical modes are combined to obtain the two output modes. The interaction of $\hat{a}_{S}$ and phase-shifted reference beam $\hat{a}_{R\phi}=\hat{a}_Re^{i\phi}$  yields modes
\begin{equation}
	\hat{a}_3=(\hat{a}_S-\hat{a}_R e^{i\phi})/\sqrt{2},\;
	\hat{a}_4=(\hat{a}_S+\hat{a}_R e^{i\phi})/\sqrt{2}.
\end{equation}
After interaction of the reference beam $\hat{a}_Re^{i\phi}$ and the output optomechanical mode $\hat{a}_S$, the resultant modes $\hat{a}_3$ and $\hat{a}_4$ are detected on the photodetectors through measured photocurrents ($I(\phi)$), which are proportional to the corresponding number operators ($\hat{I}_k(\phi)\propto \hat{n}_k (\phi)$). Following the standard continuous-variable quantum-optical measurement formalism, the Stokes-like difference operator is given by \cite{stokes1,Fabre}
\begin{equation}\label{stokedefn}
	\hat{\mathcal{S}}_1(\phi)=\hat{n}_{4}-\hat{n}_{3} ,\; \mbox{with}\; \hat{n}_{k}=\hat{a}_{k}^\dagger \hat{a}_{k},\; k=3,4.
\end{equation}
The phase-dependent photon-number difference provides access to the optical field quadratures, following the principles of balanced homodyne detection \cite{cavity2}. For  the phase-shift $\phi=0$,  we now get
\begin{equation}\label{M10}
	\langle \hat{x}_S\rangle =\frac{\langle\hat{\mathcal{S}}_1(0)\rangle}{\sqrt{2}d_r} \implies \Re(\mathcal{M}_1)= \frac{\langle\hat{\mathcal{S}}_1(0)\rangle}{2d_r},
\end{equation}
whereas, measurement for the phase-shift $\phi=\pi/2$ yields
\begin{equation}\label{M1pi2}
	\langle \hat{p}_S\rangle =\frac{\langle\hat{\mathcal{S}}_1(\pi/2)\rangle}{\sqrt{2}d_r} \implies \Im(\mathcal{M}_1)= \frac{\langle\hat{\mathcal{S}}_1(\pi/2)\rangle}{2d_r}.
\end{equation}
Combining ~\eqref{M10} and ~\eqref{M1pi2} we get the experimentally determined value
\begin{equation}
	\mathcal{M}_1^{\mathrm{expt}}= (\langle\hat{\mathcal{S}}_1(0)\rangle + i \langle\hat{\mathcal{S}}_1(\pi/2)\rangle)/(2d_r),
\end{equation}
which can be compared with the theoretical value ~\eqref{M1M2theory}. Now for the same reference beam, the second Stokes moment reads
\begin{equation}
	\langle \hat{\mathcal{S}}_1^2(\phi)= 2M_R \Re(e^{-2i\phi} \mathcal{M}_2) +n_L(2N_R+1)+N_R,
\end{equation}
where 
\begin{equation}
	N_R=d_r^2+\bar{n}_r \cosh2\theta_r +\sinh^2\theta_r,\;
	M_R=d_r^2 -(\bar{n}_r+1/2)\sinh2\theta_r.
\end{equation}
Measurements at reference angles $\phi=0,\pi/2,\pi/4$, yield following three equations
\begin{eqnarray}
	\mathscr{S}_0 &=&\langle \hat{\mathcal{S}}_1^2(0)\rangle = 2M_R\Re(\mathcal{M}_2) + n_L(2N_R+1)+N_R, \label{S1s0}\\
	\mathscr{S}_{\pi/2} &=& \langle \hat{\mathcal{S}}_1^2(\pi/2)\rangle = -2M_R\Re(\mathcal{M}_2) +n_L(2N_R+1)+N_R, \label{S1spi2}\\
	\mathscr{S}_{\pi/4} &=& \langle \hat{\mathcal{S}}_1^2(\pi/4)\rangle = 2M_R\Im(\mathcal{M}_2) +n_L(2N_R+1)+N_R. \label{Sp4}
\end{eqnarray}
Equations~\eqref{S1s0}-\eqref{Sp4} yield the experimental estimates of $n_L$ and $\mathcal{M}_2$ :
\begin{eqnarray}
	n_L &=&\frac{\mathscr{S}_{0}+\mathscr{S}_{\pi/2}-2N_R}{2(2N_R+1)},\;\;
	\mathcal{M}_2^{\mathrm{expt}}=\Re(\mathcal{M}_2^{\mathrm{expt}}) + i \Im(\mathcal{M}_2^{\mathrm{expt}}),\\
	\mbox{with}\; 
	\Re(\mathcal{M}_2^{\mathrm{expt}}) &=& \frac{\mathscr{S}_{0}-\mathscr{S}_{\pi/2}}{4M_R},\;
	\Im(\mathcal{M}_2^{\mathrm{expt}})= \frac{1}{2M_R}(\mathscr{S}_{\pi/4}-N_R-n_L(2N_R+1)).
\end{eqnarray}
Accordingly, the experimentally constructed covariance matrix is given by
\begin{equation}
	V_S^{\mathrm{expt}}=\left(\begin{array}{cc}
V_{S,11}^{\mathrm{expt}} & V_{S,12}^{\mathrm{expt}}\\
V_{S,12}^{\mathrm{expt}} & V_{S,22}^{\mathrm{expt}}
	\end{array}\right),
\end{equation}
where $V_{S,ij}^{\mathrm{expt}}$ are explicitly given by the measured photocurrents moments and calibrated reference beam parameters as follows.
\begin{eqnarray}
V_{S,11}^{\mathrm{expt}} &=& \frac{1}{2}+ n_L +  \frac{1}{4M_R}(\mathscr{S}_{0}-\mathscr{S}_{\pi/2}) - \frac{1}{2d_r^2} \langle \hat{\mathcal{S}}_1(0)\rangle^2,\\
V_{S,22}^{\mathrm{expt}} &=& \frac{1}{2}+ n_L -  \frac{1}{4M_R}(\mathscr{S}_{0}-\mathscr{S}_{\pi/2}) - \frac{1}{2d_r^2} \langle \hat{\mathcal{S}}_1(\pi/2)\rangle^2,\\
V_{S,12}^{\mathrm{expt}} &=& \frac{1}{2M_R}(\mathscr{S}_{\pi/4}-N_R-n_L(2N_R+1))  - \frac{1}{2d_r^2} \langle\hat{\mathcal{S}}_1(0)\rangle \langle \hat{\mathcal{S}}_1(\pi/2)\rangle.
\end{eqnarray}
\subsection{Determination of NC-parameters }
Now after determining $\mathcal{M}_j^{\mathrm{expt}}$ experimentally, one can determine the experimental values of $A_M$ and $\Phi_K$ through ~\eqref{RM1} and ~\eqref{RM2}. In particular, by solving the transcendental equation 
\begin{equation}
	\Phi_K+n_L\sin2\Phi_K=\arg(\mathcal{M}_1)-\arg(\alpha_l)
\end{equation}
numerically, we obtain experimental value $\Phi_K^{\mathrm{expt}}$ and use it in
\begin{equation}
	A_M=n_L(\cos2\Phi_K-1)-\ln(\vert\mathcal{M}_1/\vert\alpha_l\vert)
\end{equation} 
to determine $A_M^{\mathrm{expt}}$. The normal-mode frequencies can be determined using the Prony method for the reconstruction of finite sums of complex exponentials\cite{prony1,prony2}. We can use ~\eqref{Kerrphase} and ~\eqref{AMexpt} to write the two frequency spectral estimation problem \cite{prony1,prony2} in terms of complex experimental response $\mathscr{Z}=\ddot{A}_M+i\ddot{\Phi}_K$ as
\begin{equation}\label{spectralest}
	\mathscr{Z}(t)=\beta_{\lambda_1}e^{i\lambda_1 t} +\beta_{\lambda_2}e^{i\lambda_2 t},\; \mbox{with}\; \beta_{\lambda_j}=g^2\vert f_j\vert^2\ge 0.
\end{equation}
Here dot denotes derivative with respect to time. For equally spaced time sampling\cite{prony1,prony2} $t_k=k\Delta t$, define $\mathscr{Z}_{k}=\mathscr{Z}(t_k)$ and $z_j=e^{i\lambda_j \Delta t}$, which reduces ~\eqref{spectralest} to
\begin{equation}\label{Zk}
	\mathscr{Z}_k=\beta_{\lambda_1}z_1^k + \beta_{\lambda_2}z_2^k.
\end{equation} 
~\eqref{Zk} immediatelt implies that the complex response satisfies
\begin{equation}\label{recurence}
	\mathscr{Z}_{k+2}-(z_1+z_2)\mathscr{Z}_{k+1}+z_1z_2 \mathscr{Z}_k=0,\; \mbox{with}\; z_j=e^{i\lambda_j \Delta t}.
\end{equation}
Solving ~\eqref{recurence} we obtain $z_j$, which provides the experimental estimation of the normal mode frequencies $\lambda_j$ $(j=1,2)$, which encodes the information of NC-parameters $\theta,\eta$ explicitly. On the other hand, we have theoretical exact values of $\lambda_j$ through ~\eqref{lambdaformexplicit}. Using explicit expressions, we get the following equations of the NC-parameters $\theta$ and $\eta$.
\begin{equation}\label{thetaetaequation}
	\theta^4-\lambda_{\omega2}\theta^2+\lambda_{\omega0}=0,\;\; 
	\theta\eta=4\hbar^2\epsilon_{\theta\eta},
\end{equation}
where
\begin{eqnarray}
	\epsilon_{\theta\eta} &=&1-\sqrt{\frac{\lambda_1\lambda_2}{\tilde{\omega}_1\tilde{\omega}_2}},\;\;
	\lambda_{\omega0}=\frac{16\hbar^4\epsilon_{\theta\eta}^2}{m_1^2m_2^2 \tilde{\omega}_1^2\tilde{\omega}_2^2},\\
	\lambda_{\omega2} &=& \frac{\lambda_1^2+\lambda_2^2-(\tilde{\omega}_1^2+\tilde{\omega}_2^2)( 1+\epsilon_{\theta\eta})^2}{m_1m_2 \tilde{\omega}_1^2\tilde{\omega}_2^2}.
\end{eqnarray}
Since $\epsilon_{\theta\eta}$, $\lambda_{\omega0}$ and $\lambda_{\omega2}$ are given in terms of the bare frequencies ($\tilde{\omega}_1, \tilde{\omega}_2$) and mass ($m_1,m_2$) of the oscillator and experimentally determined normal mode frequencies $\lambda_j^{\mathrm{expt}}$, one can solve ~\eqref{thetaetaequation} and determine the values of NC-parameter $\theta,\eta$.
\subsection{Estimate of the experimental bound}
For simplicity, let us only consider the case for position-position noncommutativity (i.e., $\eta=0$). From ~\eqref{para1}-\eqref{discriminant}, we see that, for $\eta=0$, the eigenvalues $\lambda_j$ depends on $\theta^2$. Hence, all the expressions that depend on $\lambda_j(\theta^2)$ are functions of $\theta^2$. Therefore, it is reasonable to expand the Stokes-parameter in terms of $\theta^2$ as
\begin{equation}\label{s1}
	\langle \hat{\mathcal{S}}_1(\phi)\rangle_{\theta} = \langle \hat{\mathcal{S}}_1(\phi)\rangle_{\theta =0} + C_{\phi}\theta^2 + \mathcal{O}(\theta^4);\; \mbox{with}\; C_{\phi}=\frac{\partial}{\partial (\theta)^2}\langle \hat{\mathcal{S}}_1(\phi)\rangle_{\theta} \vert_{\theta=0}.
\end{equation}
On the other hand, for an ideal detector, if the measurement time is $T_d$ and $e$ is the elementary charge, we can rewrite ~\eqref{stokedefn} as
\begin{equation}\label{c1}
\langle \hat{\mathcal{S}}_1(\phi)\rangle = (T_d/e) \Delta I(\phi),\; \mbox{with}\; \Delta I(\phi)=I_4(\phi)-I_{3}(\phi)
\end{equation}
Combining ~\eqref{s1} and ~\eqref{c1}, we can write the estimated value of $\theta$ as
\begin{equation}
\Delta I_{\theta}(\phi)-\Delta I_{\theta=0}(\phi)=\delta I_{\phi}\simeq (e/T_d) C_{\phi}\theta^2\implies	\theta_{\mathrm{est}} \simeq \sqrt{\frac{T_d\delta I_{\phi}}{eC_{\phi}}}.
\end{equation}
Thus, we see that, we can actually relate the NC-parameter to the measured photocurrents and known parameters.
\section{Conclusions}
In this work, we have investigated the correlations between phase-space coordinates of a two-dimensional anisotropic harmonic oscillator in a noncommutative (NC) phase space. Through symplectic diagonalization, we have obtained the modified normal-mode frequencies and established their dependence on the NC parameters $\theta$ and $\eta$.
We have further examined the Gaussian ground state $\vert 0,0\rangle$ of the system and derived its covariance matrix in terms of the parameters characterizing  $\vert 0,0\rangle$. The resulting covariance structure reveals correlations between different canonical degrees of freedom induced by the effective coupling associated with phase-space noncommutativity. \\
An important aspect of the present investigation is the connection between NC mechanical dynamics and optical observables. By considering the interaction of the oscillator with a coherent optical field $\hat{a}_L$, we have derived analytical expressions for the optical coherence attenuation $A_M$ and the nonlinear Kerr phase $\Phi_K$. Both quantities exhibit characteristic temporal variations governed by the modified mechanical normal-mode frequencies and the corresponding optomechanical coupling strengths. Consequently, the effects of phase-space noncommutativity are encoded not only in the mechanical spectrum and ground state correlations, but also in the dynamical response of an interacting optical field. This provides a theoretical basis for probing the underlying mechanical properties through optical measurements. We have proposed a Stokes-like interferometric measurement scheme involving a phase-controlled reference beam $\hat{a}_{R\phi}$, a balanced beam splitter, and two photodetectors. The proposed arrangement provides access to phase-sensitive optical interference signals through the measured photocurrents. Within the idealized optomechanical model, the time-dependent attenuation and Kerr phase contain sufficient spectral structure to reconstruct the mechanical normal-mode frequencies and their optical coupling strengths, provided the relevant optical quantities can be independently inferred from the measured statistics. In this manner, the optical response offers a possible indirect route to characterizing the experimental values of $\theta$ and $\eta$.\\
Although changes in the normal-mode frequencies and optical response may provide signatures of noncommutative dynamics, an unambiguous determination of both parameters requires sufficient independent experimental information. In particular, the sensitivity of the optical response to the individual deformation parameters, the possible existence of multiple solutions, and the influence of experimental uncertainties must be considered when assessing the procedure in an actual experiment. These considerations are especially relevant in the weak-deformation regime, where the leading changes in the optical observables may be of second order in the noncommutative parameters. In particular, a complete experimental implementation would require an explicit optical input-output formulation incorporating cavity damping, mechanical dissipation, thermal fluctuations, and finite photodetection efficiency. Another important issue is that, in an actual experiment, the temperature should be maintained sufficiently low so that the mechanical state may be approximated by its ground state. \\
We would like to emphasize that our proposed Stokes-like measurement scheme does not rely on the assumption that spacetime must be quantized. Rather, it investigates whether deviations from the canonical phase-space algebra, if present and adequately described by the adopted deformation, can produce measurable modifications of the mechanical spectrum and the optical response.
\section{Conflict of interests}
All the authors declare that there is no conflict of interest from funding agency or any other means whatsover.  
\section{Acknowledgement} 
 P. Patra is grateful to ANRF (Formerly SERB), Govt. of India, for financial support through project grant No. EEQ/2023/000784.
We would like to thank the anonymous referee for fruitful suggestions.

\end{document}